\documentclass[12pt]{iopart}
\usepackage{graphicx, epstopdf}
\expandafter\let\csname equation*\endcsname\relax
\expandafter\let\csname endequation*\endcsname\relax
\usepackage{amsmath}
\usepackage{amssymb}
\usepackage{amstext}
\usepackage{braket}
\usepackage{array,multirow}
\usepackage{multicol}
\usepackage{booktabs}
\usepackage{adjustbox}
\usepackage{multirow}
\usepackage{capt-of}
\usepackage[backend=bibtex,style=numeric-comp,sorting=none]{biblatex}
\usepackage[utf8]{inputenc}
\usepackage{float}
\DeclareUnicodeCharacter{2212}{\textendash}

\bibliography{thebibCH.bib}
\begin{document}

\title{Cold CH radicals for laser cooling and trapping}
\author{J. C. Schnaubelt$^{\dag}$, J. C. Shaw and D. J. McCarron$^{*}$}
\address{Department of Physics, University of Connecticut, 196A Auditorium Road, Unit 3046, Storrs, CT 06269-3046}
\ead{$^{\dag}$joseph.schnaubelt@uconn.edu, $^{*}$daniel.mccarron@uconn.edu}
\vspace{10pt}

\begin{abstract}
Ultracold CH radicals promise a fruitful testbed for probing quantum-state controllable organic chemistry. In this work, we calculate CH vibrational branching ratios (VBRs) and rotational branching ratios (RBRs) with ground state mixing. We subsequently use these values to inform optical cycling proposals and consider two possible radiative cooling schemes using the $X^{2}\Pi \leftarrow A^{2}\Delta$ and $X^{2}\Pi \leftarrow B^{2}\Sigma^{-}$ transitions. As a first step towards laser cooled CH, we characterize the effective buffer gas cooling of this species and produce $\sim5\times10^{10}$ CH molecules per pulse with a rotational temperature of 2(1)~K and a translational temperature of 7(2)~K. We also determine the CH-helium collisional cross section to be $2.4(8)\times10^{-14}$~cm$^{2}$. This value is crucial to correctly account for collisional broadening and accurately extract the in-cell CH density. These cold CH molecules mark an ideal starting point for future laser cooling and trapping experiments and tests of cold organic chemistry.    

\end{abstract}

\section{Introduction}
Ultracold molecules provide a versatile platform for quantum simulations, tests of fundamental symmetries, and quantum chemistry \cite{Carr2009}. Over the past decade, much success has been found using molecules associated from pairs of pre-cooled atoms \cite{Ni08}. Naturally, this approach limits users to species made from laser-coolable atoms and, for a broader range of molecules, direct cooling techniques must be employed. To date, a variety of diatomic and polyatomic molecules, well-suited to several proposed applications, have been directly cooled \cite{Prehn2016,McCarron2018, Mitra2020}. A subset of these molecules has been successfully laser cooled and trapped in 3D magneto-optical traps (MOTs), providing a robust starting point for studies of ultracold molecular physics. Today, trapped samples of laser-cooled molecules are limited to metal-fluorides \cite{Barry2014, Anderegg2017, Williams2017} and metal-oxides \cite{Collopy2018} chosen for their favorable properties and experimental convenience. We plan to draw on the success of these experiments and apply laser cooling and trapping techniques to a more chemically relevant molecule, methylidyne (CH). Future research using this molecule may include tunable organic chemistry and improved precision measurements \cite{Koslov2009, Beyer2017}.

Ultracold samples of laser-cooled CH will broaden the already rich science surrounding this molecule both in and out of the laboratory. CH was the first molecule discovered in the interstellar medium (ISM) \cite{Douglas1941} and remains a topic of intense interest regarding the formation of larger hydrocarbons in interstellar molecular clouds \cite{Cooke2019}. Low temperature reaction rates and kinetic measurements of interstellar CH have yet to be reproduced in the laboratory, and could be used to guide searches for astrochemical reactions \cite{Smith2011}. In addition to extensive studies in the ISM, CH has previously been studied in crossed molecular beam experiments \cite{Maksyutenko2011}, sensitive searches for variations of fundamental constants \cite{Truppe2013}, and cold methyl molecules (CH$_{3}$), closely related to CH, have been magnetically trapped to probe cold (200~mK) chemistry \cite{Liu2017}. Trapped samples of ultracold CH may enable precise measurements of the formation of neutral acetylene and the higher energy isomer, vinylidene \cite{DeVine2017}. Additionally, the careful dissociation of CH may provide a route to ultracold carbon and hydrogen that avoids the experimental challenges of laser cooling these atoms directly \cite{Kielpinski2006,Burkley19}.

The chemical relevance of CH comes at the cost of experimental complexity stemming from its rovibrational structure. Fortunately, large recoil velocities reduce the number of scattered photons required to slow and cool a beam of molecules, as addressed in the following section. CH is also amenable to electrostatic guiding and Stark deceleration due to its low mass, $\Lambda$-doublet structure, and substantial dipole moment (1.46~D)\cite{TruppeThesis}. Stimulated optical forces also appear to be promising due to its large recoil velocities and relatively long-lived excited states \cite{Chieda2011}. 

In this work, we consider the feasibility of laser cooling CH by calculating vibrational and rotational branching ratios for two electronic transitions. We then propose two optical cycling schemes to tackle this branching with a level of complexity similar to those used in current experiments with other species \cite{Baum2021}.  Finally, as a first step towards laser cooling CH, we demonstrate effective buffer gas cooling of this species and present measured properties of our cold CH radicals.

\section{Theory}

\subsection{Electronic Structure}
Upon first inspection, there are four excited electronic states of interest in CH: $a^{4}\Sigma^{-}$, $A^{2}\Delta$, $B^{2}\Sigma^{-}$, and $C^{2}\Sigma^{+}$ \cite{Lane2011, Umbachs1986, Yarkony1994}. For effective laser cooling, it is necessary to choose an electronic transition with highly diagonal Franck Condon Factors (FCFs), limited rotational branching, and a short-lived excited state to ensure rapid optical cycling. While the $C^{2}\Sigma^{+}$ state has diagonal FCFs, it also has a large predissociation probability making it poorly suited for optical cycling \cite{Lane2011,Lutz1970}. Similarly, we do not consider the $a^{4}\Sigma^{-}$ excited state due to its long lifetime, estimated to be up to $\sim$12 seconds \cite{Yarkony1994}. The low mass of CH (13~amu), and electronic transitions in the blue (430~nm) or near-ultraviolet (389~nm), lead to large recoil velocities of 7.1~cm~s$^{-1}$ and 7.9~cm~s$^{-1}$ for the remaining $X^{2}\Pi \leftarrow A^{2}\Delta$ and $X^{2}\Pi \leftarrow B^{2}\Sigma^{-}$ cycling transitions, respectively. By comparison, these are both $>$12$\times$ greater than the recoil velocity of SrF (0.56~cm~s$^{-1}$), the first molecule to be laser-cooled and trapped \cite{Shuman2010, Barry2014}. This increased recoil velocity greatly reduces the number of photons required to laser cool and trap CH and therefore the required degree of rovibrational closure. This benefit of CH is especially relevant since the ground electronic state, $X^{2}\Pi$ ($\Lambda''$=1), is split into opposite parity $\Lambda$-doublets, eliminating the convenient use of parity to confine rotational branching \cite{Stuhl2008}. (In this work $v$, $N$, $S$, and $\Lambda$ are used to define the vibrational, rotational, spin, and axial orbital angular momentum quantum numbers respectively, with the ground state denoted by ($''$) and excited state by ($'$)). Lastly, as we discuss in later sections, the $A^{2}\Delta$ state has a minimum rotational state of $N'$=2 (since $N \geq \Lambda$), making addressing every rotational branch more challenging.

\subsection{Vibrational Structure}
Substantial previous work on CH provides many sources for spectroscopic constants \cite{ZACHWIEJA1995, KEPA1996, Lane2011, Masseron2014, Kalemos1999, BERNATH1991, Cui2018, NEMES1999} that can be used to generate the electronic potentials and vibrational wavefunctions needed to calculate vibrational branching ratios (VBRs). Our approach uses the Le Roy suite of programs to first calculate potential energy curves from vibrational and rotational constants with the Rydberg Klein Rees (RKR) method to calculate classical turning points \cite{LEROY2017}. Potential energy curves, along with the relevant electronic offsets ($T_{e}$-values), are then used with LEVEL \cite{LEROY2007}, which numerically solves the 1D Schrodinger equation to calculate vibrational wavefunctions ($\psi_{v}$), as illustrated in figure 1. Finally, FCFs are calculated using the overlap integral $|\braket{\psi_{v'}|\psi_{v''}}|^{2}$.

Here, we present calculated VBRs from the $A^{2}\Delta$, and $B^{2}\Sigma^{-}$ excited states to the $X^{2}\Pi$ ground state. A range of constants from a variety of spectroscopic data sets are included for each excited state, allowing us to quantify an uncertainty through multiple calculations of the VBRs.

The VBRs are calculated using
\begin{equation}
    {\rm{VBR}}_{v',v''}=\frac{q_{v',v''} \times \omega_{v',v''}^{3}}{\sum_{i=0}^{\infty}q_{v',i}\times\omega_{v',i}^{3}},
\end{equation}
where $\omega_{v',v''}$ is the energy difference between the ground ($v''$) and excited ($v'$) vibrational states under consideration, and $q_{v',v''}$ is the relevant FCF \cite{Barry2013}. Here, each FCF is weighed by its respective energy separation ($\omega_{v',v''}$) to the third power, leading to calculated VBRs from $v'=0$ being more diagonal compared to the corresponding FCFs. Additionally, calculated VBRs from excited vibrational states ($v'>0$) are biased towards decays with $v''<v'$ compared to the FCFs. Both of these changes limit branching into more highly excited vibrational states, which improves prospects for applying typical laser cooling schemes. The difference between FCFs and VBRs can be especially pronounced for molecules with large vibrational constants ($\omega_{e}$) relative to the transition electronic offset ($T_{e}$), such as BH \cite{Hendricks2014}, CH, and CaH \cite{DiRosa2004}.

Our calculated VBRs, with uncertainties stated in parentheses, are listed for the $X^{2}\Pi \leftarrow A^{2}\Delta$ and $X^{2}\Pi \leftarrow B^{2}\Sigma^{-}$ transitions in table 1 and agree with previous calculations \cite{Lane2011, ZACHWIEJA1995, KEPA1996}. Figure 1 shows the vibrational wavefunctions that were used to calculate the first three VBRs, highlighting the strong anharmonicity of the $B^{2}\Sigma^{-}$ state, relative to the $X^{2}\Pi$ state, which results in substantial off-diagonal FCFs.

\begin{table}[H]
\begin{adjustbox}{center, width=\linewidth}
\begin{tabular}{lcccc} \\ \hline \multicolumn{5}{c}{$X^{2}\Pi \leftarrow A^{2}\Delta$} \\ \hline \hline 
$v' \setminus v''$ & 0 & 1 & 2 & $\geq$3  \\ \hline 
\multicolumn{1}{l}{0}    & 0.9940(3)   & 0.0058(3)  & 0.00012(2)  &  \multicolumn{1}{l}{$8\times 10^{-5}$(5)} \\
\multicolumn{1}{l}{1}   &  0.0110(4) &  0.981(1) & 0.0075(8)  &  \multicolumn{1}{l}{0.0005(2)} \\
\multicolumn{1}{l}{2}    &  0.00115(8) &  0.0127(1) & 0.982(3)  &  \multicolumn{1}{l}{0.0042(2)}  \\
\multicolumn{1}{l}{3}    &  $6\times 10^{-5}(2)$ &  0.0037(6) & 0.007(3)  & \multicolumn{1}{l}{0.989(3)} \\ \hline \hline 
\multicolumn{5}{c}{$X^{2}\Pi \leftarrow B^{2}\Sigma^{-}$}\\ \hline \hline $v' \setminus v''$ & 0 & 1 & 2 & $\geq$3 \\ \hline  \multicolumn{1}{l}{0}    & 0.902(4)   & 0.090(3)  & 0.0076(6)  &  0.00040(7) \\ \multicolumn{1}{l}{1}    &  0.168(9) &  0.55(1) & 0.231(3)  &  0.0510(3) \\  \multicolumn{1}{l}{2}    &  0.051(2) &  0.57(1) & 0.13(1)  &  0.249(5) \\ \multicolumn{1}{l}{3}    &  0.0082(4) &  0.417(6) & 0.54(1)  & 0.034(7) \\ \hline 
\end{tabular}
\end{adjustbox}

\caption{Calculated VBRs for the $X^{2}\Pi \leftarrow A^{2}\Delta$ and $X^{2}\Pi \leftarrow B^{2}\Sigma^{-}$ transitions. Uncertainty, stated in the last decimal place, encompasses the range of VBR values obtained using constants from several different sources \cite{ZACHWIEJA1995, KEPA1996, Lane2011, Masseron2014, Kalemos1999, BERNATH1991, Cui2018, NEMES1999}.}
\end{table}

\begin{figure}[H]
\centering
\includegraphics[width=\linewidth]{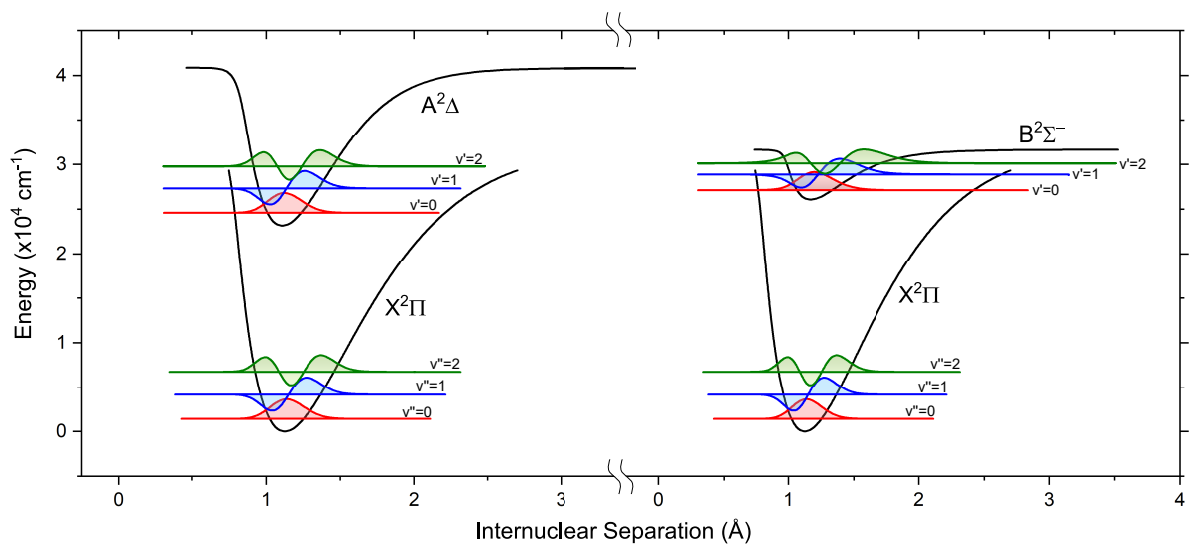}
\caption{Electronic potentials with vibrational wavefunctions used to calculate the VBRs shown in table 1. Differing levels of anharmonicity in the $B^{2}\Sigma^{-}$ and $X^{2}\Pi$ states result in substantial off-diagonal FCFs and large vibrational branching.}
\end{figure}

\subsection{Rotational Structure}
A critical potential loss channel in CH is the complicated rotational structure of the ground $X^{2}\Pi$ state owing to a nonzero orbital angular momentum along the internuclear axis. This results in $\Lambda$-doublets for each $J''$ (where $J=N\pm S$) which eliminate the possibility of using parity to suppress rotational branching through selection rules \cite{Stuhl2008}, making CH rotationally similar to many polyatomic species \cite{Augenbraun2020}. An appropriate description for the rotational structure requires identifying the correct angular momentum coupling scheme for each electronic state. Due to its small molecular mass, the electronic angular momentum is coupled to the internuclear axis, and thus CH is not described well by Hund's case (c), (d), or (e) basis states \cite{Brown2003}. 

A good description of rotational states by Hund's case (a) or (b) occurs when there is a large energy difference between rotational and spin-orbit splitting, $A\Lambda\gg BJ$ for (a) and $A\Lambda\ll BJ$ for (b) \cite{Field04}. Here $A$ is the spin-orbit coupling constant and $B$ is the rotational constant. For the CH $X^{2}\Pi$ ground state, where $B$=14.457~cm$^{-1}$, $A$=28.1~cm$^{-1} $\cite{TruppeThesis} and $\Lambda''=1$, the ratio of $A/B$ $\simeq 2$, indicating that at low energies, CH is not described well by pure basis states and is only described well in Hund's case (b) in higher rotational states ($J''\gg 2$). As a result, we choose to describe CH in Hund’s case (b), but note that due to the large spin-orbit coupling, $N''$ is not a good quantum number for the lower $J''$ states relevant to laser cooling and trapping experiments.

A ground state Hamiltonian is constructed to calculate mixing between spin-orbit states, and enable predictions of transition frequencies. We construct an effective Hamiltonian as 

\begin{equation}
    H_{\rm{effective}}=H_{\rm{rotation}}+H_{\rm{spin-orbit}}+H_{\rm{spin- rotation}}+H_{\rm{hyperfine}},
\end{equation}
following the example in section 10.6.3 of \cite{Brown2003}. The complete Hamiltonian is diagonalized with a matrix of possible Hund’s case (b) basis states, yielding eigenvalues describing the ground state energy levels. Corresponding eigenvectors indicate a mixing between states with common $J''$ due to a large spin-orbit coupling constant. For the $X^{2}\Pi \leftarrow B^{2}\Sigma^{-}$ transition, this mixing leads to rotational branching (discussed below) from the $\ket{v'=0; N'=0, J'=1/2}$ state into the $\ket{v''=0; N''=2, J''=3/2}$ state, of about $4.4\%$, which would be forbidden if this mixing were neglected, and $N$ was treated as a good quantum number. 

Once we have quantified mixing between adjacent states it is crucial that we understand the rotational branching ratios (RBRs) from each excited state. We calculate RBRs using transition matrix elements written directly in a Hund's case (b) basis, allowing for the prediction of branching ratios \cite{Willitsch2018, Wall2009}. During a spontaneous emission event, the probability of populating a certain ground state is determined by the RBRs combined with the mixing calculated using the ground state Hamiltonian. These probabilities are shown in parentheses within figure 2, where we have excluded $\Lambda$-doubling and hyperfine structure for simplicity. 

\begin{figure}[H]
\centering
\includegraphics[width=4in]{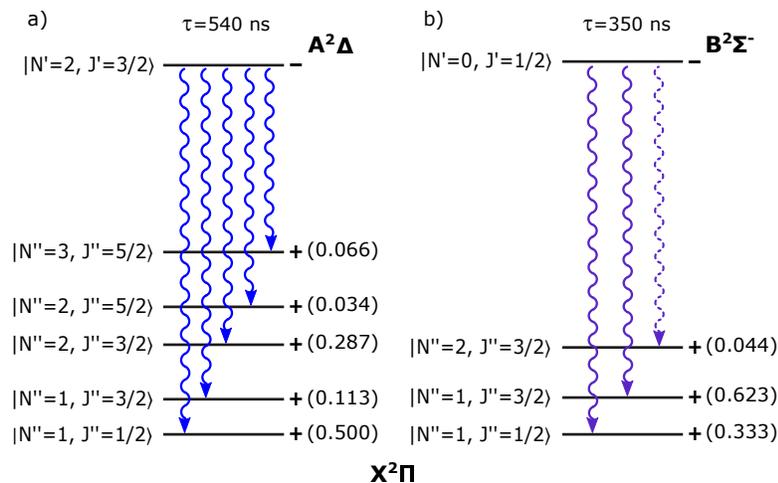}
\caption{Rotational branching for the two optical cycling transitions considered only showing even parity ground states and neglecting hyperfine structure. (a) The $X^{2}\Pi \leftarrow A^{2}\Delta$ cycling transition does not allow decay into $J''=7/2$, which fortunately means that no additional decay pathways are introduced by mixing. (b) Mixing between $N''=1$ and $N''=2$ provides a potentially critical loss channel for the $X^{2}\Pi \leftarrow B^{2}\Sigma^{-}$ transition, as discussed in the text.}
\end{figure}

\subsection{Radiative Cooling Approach}

\subsubsection{$X^{2}\Pi \leftarrow A^{2}\Delta$}
Rotational branching means that cooling on the $X^{2}\Pi \leftarrow A^{2}\Delta$ transition requires five lasers per vibrational level to ensure rotational closure (figure 2(a)). Neglecting mixing, the RBRs into the ground rotational states are, in ascending energy, $1/2:1/10:3/10:1/30:1/15$. Including mixing alters these values as shown in figure 2(a). The large number of rotational states populated presents a challenge to optical cycling by decreasing the achievable scattering rate on the $X^{2}\Pi \leftarrow A^{2}\Delta$ transition, governed by equation 3 below.

\subsubsection{$X^{2}\Pi \leftarrow B^{2}\Sigma^{-}$}
The $X^{2}\Pi \leftarrow B^{2}\Sigma^{-}$ transition features more favorable rotational branching than the $X^{2}\Pi \leftarrow A^{2}\Delta$ transition, countering the drawbacks of the off-diagonal VBRs (see table 1). Upon first inspection, there are only 2 possible decay pathways from the $\ket{N'=0, J'=1/2}$ excited state. However, ground state mixing results in a $\Delta N=2$ transition, typically forbidden for Hund's case (b) molecules, and leads to 3 possible decay pathways, and a need for three lasers per vibrational state (figure 2(b)). There is also an added risk of dissociation for molecules that are repumped into the $B^{2}\Sigma^{-}$ state; previous work shows that predissociation from the first excited vibrational state in low rotational states is possible \cite{Erman1976, vanDishoeck1987}.

\subsubsection{Hyperfine Structure}
Each $X^{2}\Pi$ spin-rotation state is split into two hyperfine sub-levels, due to the spin of the hydrogen nucleus ($I=1/2$). This splitting is on the order of 10 MHz and can be easily spanned by common modulation techniques. This is in contrast to spin-rotation structure, which spans $\sim$530~GHz and $\sim$190~GHz in $N''=1$ and $N''=2$ respectively, and thus demands multiple lasers \cite{Truppe2013}. To prevent population loss into dark Zeeman sublevels, one can use polarization modulation for all rovibrational states \cite{Berkeland2002}. One can also remix dark sublevels through Larmor precession in those specific ground states with nonzero Land\'{e} g-factors, where electron spin and orbital angular momentum are aligned \cite{Kloter2008}.

\subsubsection{Laser Cooling Proposal}
Laser cooling and trapping CH radicals from a cryogenic source requires $>$10$^{3}$ scattered photons. Our calculations show that sufficient rovibrational closure is possible in CH by optically addressing the lowest two and three ground state vibrational levels for the $X^{2}\Pi \leftarrow A^{2}\Delta$ and $X^{2}\Pi \leftarrow B^{2}\Sigma^{-}$ transitions, respectively (figure 3). Note that rotational branching due to spin-orbit mixing on the $X^{2}\Pi \leftarrow B^{2}\Sigma^{-}$ transition makes the excited $B^{2}\Sigma^{-}\ket{v'=0; N'=1, J'=0}$ state available for repump lasers to address without further rotational branching. This approach avoids the large and undesirable off-diagonal FCFs out of $B^{2}\Sigma^{-} \ket{v'=1; N'=0, J'=0}$, as seen in figure 1 and table 1.

While laser diodes are available to excite all of the required transitions in both proposed systems, laser cooling and trapping CH using the $X^{2}\Pi \leftarrow A^{2}\Delta$ transition is advantageous due to the availability of low-cost laser diodes at 430~nm and 488~nm. These lasers could also be repurposed to cool on the $X^{2}\Pi \leftarrow B^{2}\Sigma$ transition in the future, with the addition of three 389~nm lasers. This route may become more attractive as ultraviolet laser diodes continue to mature.

Optical cycling experiments using the $X^{2}\Pi \leftarrow A^{2}\Delta$ transition in CH would require three lasers to scatter $\simeq10$ photons (equivalent to >100 photons in SrF), sufficient to observe radiative deflection of a molecular beam. Six lasers ($\simeq300$ photons) would enable transverse cooling, while nine lasers are required to scatter $\simeq3000$ photons before $1/e$ of the original population remains. This should be adequate to slow, capture, and cool CH radicals in a MOT, where a tenth laser at 488~nm, addressing the  $X^{2}\Pi \ket{v''=1; N''=2, J''=5/2} \leftarrow A^{2}\Delta \ket{v'=0; N'=2, J'=3/2}$ transition, may be required (enabling $\simeq8000$ scattered photons) to increase the trap lifetime. Alternatively, accepting greater loss and less than $1/e$ of the original population to realize smaller trapped samples with fewer lasers may also be possible \cite{Baum2021}.

\begin{figure}[H]
\centering
\includegraphics[width=5in]{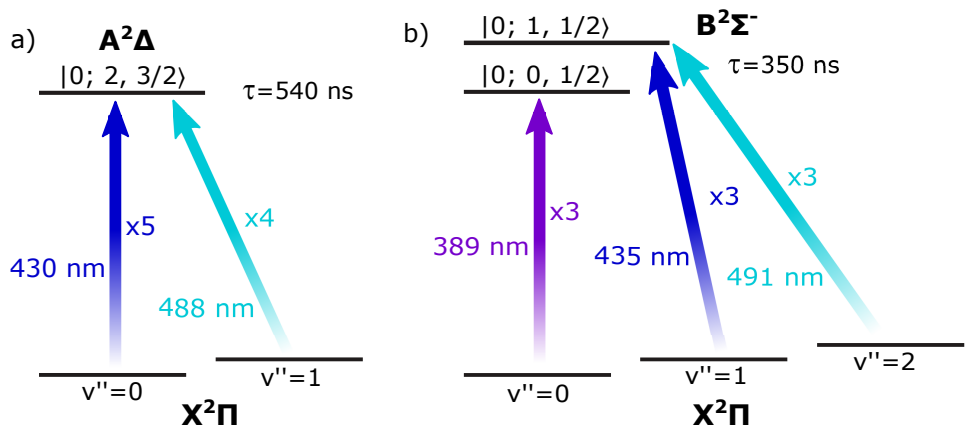}
\caption{Proposed laser cooling schemes for the (a) $X^{2}\Pi \leftarrow A^{2}\Delta$ and (b) $X^{2}\Pi \leftarrow B^{2}\Sigma^{-}$ transitions providing enough scattered photons to
laser cool and trap a beam of CH from a cryogenic source. The number of lasers addressing each vibrational transition is highlighted and excited state kets are labeled $\ket{v'; N', J'}$. }
\end{figure}

\subsubsection{Scattering Rate}
The $A^{2}\Delta$ excited state considered for laser cooling CH possesses a relatively long radiative lifetime ($\tau$=540~ns)  \cite{TruppeThesis} which can lead to modest scattering rates. The maximum photon scattering rate ($R^{\rm{max}}_{sc}$) is given by,
\begin{equation}
    R^{\rm{max}}_{sc}=\frac{n_{e}}{n_{e}+n_{g}}\Gamma,
\end{equation}
where $\Gamma$ is the spontaneous emission rate ($1/\tau$), and $n_{g}$ and $n_{e}$ are the number of ground and excited state sublevels respectively \cite{Shuman2009}. Accounting for magnetic hyperfine structure, the $X^{2}\Pi \leftarrow A^{2}\Delta$ transition has a maximum scattering rate of $R^{\rm{max}}_{sc}=[8/(76+8)](1/\tau)=1.8\times10^{5}~\rm{s}^{-1}$ for the proposed 9 laser optical cycling scheme. We note that this scattering rate is $\sim10 \times$ less than typical photon scattering rates in molecules, however the CH recoil velocity is $\sim10 \times$ larger, leading to comparable optical forces.

\section{Experimental Results}
As a first step towards laser cooling and trapping CH we have produced cold samples using a cryogenic buffer gas beam source, as described in \cite{Shaw2020}. Molecules are produced via ablation using $\sim$50~mJ pulses at 532~nm from a Nd:YAG laser focused onto a solid precursor. Once produced, elastic collisions with a helium buffer gas at $\sim$2.4~K rapidly cool the molecules. Absorption measurements are made in-cell using a homebuilt tunable external cavity diode laser (ECDL) probing the $X^{2}\Pi \leftarrow A^{2}\Delta$ transition at 430~nm. To characterize the number of molecules produced we typically probe the $\ket{v''=0; N''=1, J''=1/2^{+}} \leftarrow \ket{v'=0; N'=2, J'=3/2^{-}}$ transition with $\sim$100~$\mu$W of light. Iodoform (CHI$_{3}$), paraffin wax, and Kapton were tested as potential ablation materials. We detected no CH from Kapton, despite the success of previous work \cite{Lancaster2020}. However, iodoform and paraffin wax both produced cold CH, with iodoform producing $\sim2 \times$ more molecules with longer lived ablation spots.

\begin{figure}[H]
\centering
\includegraphics[width=4in]{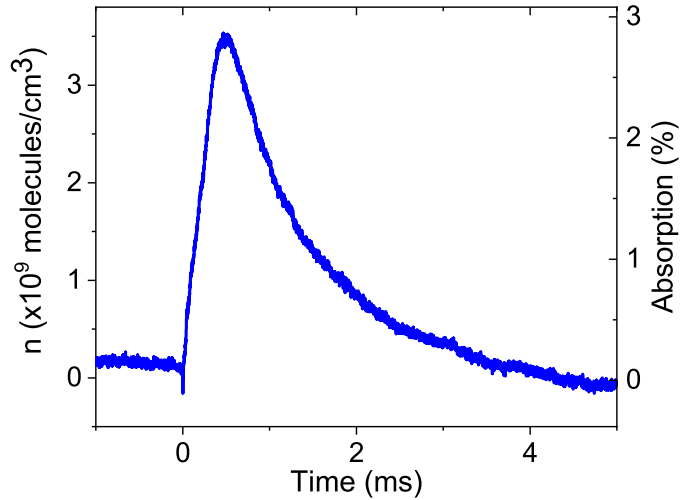}
\caption{A typical in-cell absorption trace at a temperature of $\sim$2.4~K and a helium density of 3$\times$10$^{16}$~cm$^{-3}$ shows a peak CH density of $\simeq 3.5\times 10^{9}$ cm$^{-3}$ per pulse. Less than $1/e$ of the initial population remains in the cell after 2~ms, which is substantially shorter than other species using the same cell geometry \cite{Barry2011}.}
\end{figure}

\subsection{Collsional Cross Section}
The long lifetime of the CH $A^{2}\Delta$ state (540~ns) \cite{TruppeThesis}, is shortened due to in-cell collisions with the helium buffer gas, which, when neglected could lead to a significant overestimate of the in-cell molecular density. This is typically not a problem for many of today's experiments which use either low helium densities or feature molecules with short-lived excited states ($\lesssim 100$~ns). Collisional broadening is measured by recording absorption signals (figure 4) for helium densities ranging from 10$^{16}$ to 10$^{17}$~cm$^{-3}$. The decay of these traces is fit with an exponential to obtain a time constant ($\tau$), which is the parallel addition of the time for the molecules to exit the cell ($\tau_{p}$), and the time for the molecules to diffuse from the probe beam region ($\tau_{d}$), $1/\tau = 1/\tau_{p}+1/\tau_{d}$. Following the method in \cite{Hutzler2011b}, $\tau_{p}$ is determined from our cell geometry, and helium thermal velocity. We note that less than $1/e$ of the CH peak density remains in-cell after 2~ms, which $\sim$10$\times$ shorter than SrF in the same cell geometry \cite{Barry2011}. This is similar to the short in-cell lifetimes observed in other experiments with CH \cite{Lancaster2020, Fabrikant2014}, and could be due to in part cold organic chemistry. Using these measurements, and the equation,
\begin{equation}
    \sigma_{bs}=\frac{9 \pi v_{th}\tau_{d}}{16 A_{\rm{probe}}n_{b}},
\end{equation}
we calculate the CH-helium collisional cross section ($\sigma_{bs}$) to be $2.4(8)\times10^{-14}$~cm$^{2}$, where $n_{b}$ is the in-cell helium density, $\tau_{d}$ is the extracted diffusion time, $v_{th}$ is the thermal velocity (using the CH-helium reduced mass), and  $A_{\rm{probe}}$ is the cross sectional area of the probe beam \cite{Hutzler2011b}.

\subsection{Doppler Width}
Absorption features are measured as a function of laser frequency using the $X^{2}\Pi \ket{v''=0; N''=1, J''=1/2} \leftarrow A^{2}\Delta \ket{v'=0; N'=2, J'=3/2}$ transition to obtain an in-cell linewidth. From this value, accounting for hyperfine structure and neglecting the effects of collisional broadening ($\sim$10~MHz), we determine a Doppler linewidth of 350(40)~MHz. This corresponds to an in-cell translational temperature of 7(2)~K, which is greater than our cell temperature ($\sim$2.4~K). This discrepancy could be due to heating of the buffer gas in the ablation region as reported in other experiments \cite{Barry2011}. This is especially relevant due to the relatively high YAG pulse energies used in this experiment ($\sim$50~mJ).

\subsection{Molecule Number}
The collisional lifetime, required for an accurate estimate of in-cell molecule density, can be determined with, 
\begin{equation}
\tau_{c}=\left( n_{b}\sigma_{bs}v_{th}\sqrt{1+\frac{m_{He}}{m_{CH}}}\right)^{-1},
\end{equation}
where $v_{th}$ is the helium mean thermal velocity, and assuming every collision causes an excited molecule to decay \cite{Barry2011}. Using equation 5, we determine an in-cell collisional lifetime ranging from 150 to 15~ns for helium densities from 10$^{16}$ to 10$^{17}$~cm$^{-3}$, respectively. This is substantially less than the radiative lifetime of the CH $A^{2}\Delta$ state and must be accounted for in molecule number calculations.

We typically measure $\sim$3$\%$ absorption in-cell, corresponding to a 10$\times$ larger optical density than previously reported \cite{Lancaster2020, Fabrikant2014}. The CH density is determined using Beer's Law, $I=I_{o}e^{-\sigma_{a} n l}$, where $l$ is the length of the probe region, $n$ is the density, and $\sigma_{a}$ is the absorption cross section calculated with,
\begin{equation}
	    \sigma_{a}=\frac{\lambda^{3}}{8\pi^{2}\tau_{c}}\frac{2J'+1}{2J''+1}\sqrt{\frac{m\pi}{2kT}}\times \rm{VBR} \times \rm{RBR}.
\end{equation}
In these measurements, $J'=3/2$ and $J''=1/2$ are the excited and ground state spin-rotation quantum numbers respectively, $T$ is the translational temperature, and $\rm{VBR}=0.9940$ and $\rm{RBR}=0.5$ are the vibrational and rotational branching ratios for the transition probed \cite{Budker2008}. As shown in equation 6, on the $X^{2}\Pi \leftarrow A^{2}\Delta$ transition, CH has a small absorption cross section due to its short wavelength, long excited state lifetime despite collisional broadening ($\tau_{c}$=80~ns with a helium density of 3$\times$10$^{16}$~cm$^{-3}$), low mass, and significant rotational branching. Accounting for these effects, we produce a peak density of $\simeq 3.5\times 10^{9}$ cm$^{-3}$ per pulse in the $\ket{v''=0; N''=1, J''=1/2^{+}}$ state. Assuming uniform in-cell density, this corresponds to a total molecule number of $5\times 10^{10}$ per pulse. 

\subsection{Rotational Temperature}
CH has a small moment of inertia due to its small mass (13~amu) and short bond length ($\sim$0.112~nm) \cite{ZACHWIEJA1995}, leading to a large rotational constant ($B$=14.457 cm$^{-1}$), typical of less massive hydrocarbons. This, in concert with the large spin-orbit coupling constant leads to large energy splittings between adjacent $J''$ and $N''$ levels and thus few states are populated by the Boltzmann energy distribution at cryogenic temperatures. The in-cell CH rotational temperature was measured using absorption to probe the relative ground state populations in the $\ket{v''=0; N''=1, J''=1/2}$ and $\ket{v''=0; N''=1, J''=3/2}$ $\Lambda$-doublets, as seen in figure 5. We note that the $\simeq700$~MHz $\Lambda$-doublet splitting in the $\ket{v''=0; N''=1, J''=3/2}$ state was not resolved and very few molecules were detected in these states. The in-cell CH rotational temperature is 2(1)~K, in good agreement with our cell temperature of 2.35(1)~K. To confirm resonant excitation of the sparsely populated $\ket{v''=0; N''=1, J''=3/2}$ $\Lambda$-doublet the rotational temperature measurement was repeated at an increased cell temperature of 50(5)~K. In this case, $\sim25~\%$ of the molecules populated these higher energy states and the extracted CH rotational temperature was 42(5)~K.

\begin{figure}[H]
\centering
\includegraphics[width=5in]{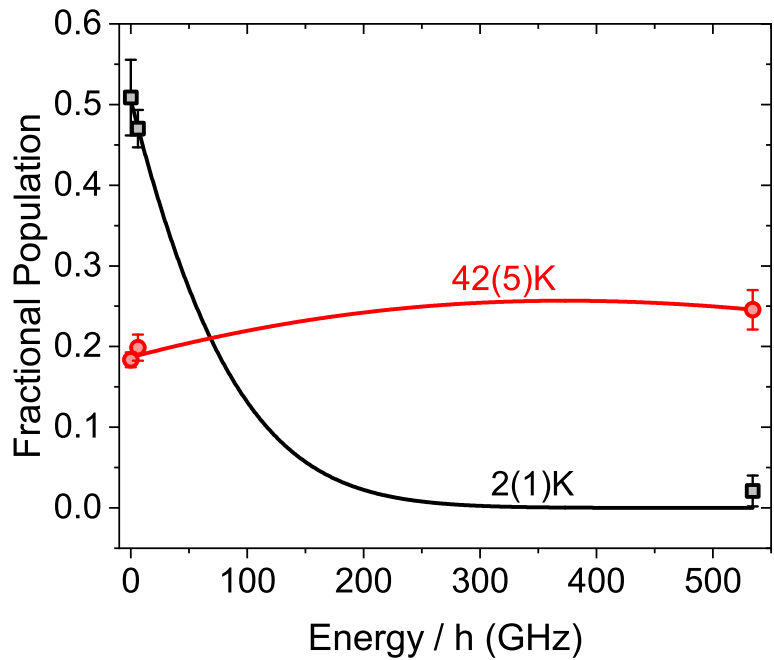}
\caption{Fractional CH populations versus state energy measured for cell temperatures of 2.35(1)~K (black squares) and 50(5)~K (red circles). Fits to the data are Maxwell-Boltzmann distributions and are used to determine the rotational temperatures shown.}
\end{figure}

\section{Conclusion}
We have calculated VBRs for CH, in good agreement with the work of others \cite{Lane2011,ZACHWIEJA1995,KEPA1996}, and RBRs which include the effects of ground state mixing. Our proposed laser cooling schemes provide sufficient rovibrational closure to slow, capture, and cool molecules in a MOT with a level of complexity comparable to those used in other experiments with other species \cite{Baum2021}. While substantial rotational branching in CH will lead to low maximum photon scattering rates compared to species laser cooled to-date, its large recoil velocity will result in comparable maximum radiative forces. CH is also amenable to stimulated optical forces, electrostatic guiding, Stark deceleration and other `few-photon’ techniques \cite{Augenbraun_2020, Fitch2016, Augenbraun2021} which may be particularly useful for beam slowing prior to loading a MOT. 

Traditionally, the production of CH has been challenging due to the highly reactive nature of this tri-radical \cite{TruppeThesis}. Despite this precedent, we have shown effective buffer gas cooling of CH radicals at densities of up to $\simeq 3.5\times10^{9}$~cm$^{-3}$. A natural extension of this work is to detect a beam of cold CH radicals exiting our cryogenic source. Assuming $\sim10~\%$ extraction \cite{Hutzler2011b} and a lack of collisional broadening, we expect absorption signals to be at the $10^{-5}$ level and beyond our current sensitivity. Fluorescence detection is an attractive alternative for beam detection and was recently used to detect an electrostatically guided beam of CH \cite{Lancaster2020}.

We typically produce $\sim5\times10^{10}$ CH molecules per pulse with a translational temperature of 7(2)~K and a rotational temperature of 2(1)~K. We have also determined the CH-helium collisional cross section to be $2.4(8) \times 10^{-14}$~cm$^{2}$. This value is crucial to properly account for the effects of collisional broadening, and thus calculate an accurate in-cell molecular density. These cold CH radicals are an ideal starting point for the production of a molecular beam for tests of our proposed optical cycling scheme and future laser cooling and trapping experiments. The rich physics accessible with cold CH may already be apparent in the relatively short $\sim$1~ms measured in-cell lifetime, suggestive of cold organic chemistry. This could potentially be probed using the ion spectrometry techniques recently extended to ultracold molecules \cite{Liu2020}.   

\section{Acknowledgements}
We thank J. D. Weinstein for helpful discussions and the loan of a 430~nm laser diode and thank P. L. Gould for his feedback on the manuscript. We acknowledge financial support from the University of Connecticut, including a Research Excellence Award from the Office of the Vice President for Research.

\printbibliography

@Article{Barry2014,
  title = {Magneto-optical trapping of a diatomic molecule},
  author = {Barry, J.~F. and McCarron, D.~J. and Norrgard, E.~N. and Steinecker, M.~H. and DeMille, D.},
  journal = {Nature},
  volume = {512},
  number = {286},
  pages = {286},
  numpages = {4},
  year = {2014},
  doi = {10.1038/nature13634},
  publisher = {Nature Publishing Group, a division of Macmillan Publishers Limited.}
}

@ARTICLE{Hendricks2014,
   author = {Hendricks, R.~J. and Holland, D.~A. and Truppe, S. and Sauer, B.~E. and Tarbutt, M.~R.},
    title = "{Vibrational branching ratios and hyperfine structure of $^{11}$BH and its suitability for laser cooling}",
  journal = {ArXiv e-prints},
archivePrefix = "arXiv",
   eprint = {1404.6174},
 primaryClass = "physics.atom-ph",
     year = 2014,
    month = Apr}

@PhdThesis{Barry2013,
           title = "{Laser cooling and slowing of a diatomic molecule}",
          author = "John F. Barry",
          school = "Yale University",
            year = "2013",
            
}

@article{Truppe2013,
author ={Truppe, S. and Hendricks, R. and Tokunaga, S and Lewandowski, H.J. Kozlov,  M.G. and Henkel, C. and Hinds, E.A. and Tarbutt M.R.},
title = {A search for varying fundamental constants using hertz-level frequency measurements of cold CH molecules},
journal = {Nat. Commun.},
volume = {4},
number = {2600},
year = {2013}
}

@article{LeRoy2007,
title = "Level 8.0: A computer program for solving the radial Schr{\"{o}}dinger equation for bound and quasibound levels",
journal = "University of Waterloo Chemical Physics Research Report CP-663",
year = "2007",
author = "Le Roy, R. J.",
}

@ARTICLE{Wall2009,
  title = "{Lifetime of the  $A(v'=0)$  state and Franck-Condon factor of the  $A-X(0-0)$  transition of CaF measured by the saturation of laser-induced fluorescence}",
  author = {Wall, T. E. and Kanem, J. F. and Hudson, J. J. and Sauer, B. E. and Cho, D.  and Boshier, M. G. and Hinds, E. A. and Tarbutt, M. R.},
  journal = {Phys. Rev. A},
  volume = {78},
  number = {6},
  pages = {062509},
  numpages = {10},
  year = {2008},
  month = dec,
  doi = {10.1103/PhysRevA.78.062509},
  publisher = {American Physical Society}
}

@ARTICLE{Shuman2009,
   author = {{Shuman}, E.~S. and {Barry}, J.~F. and {Glenn}, D.~R. and {DeMille}, D.
	},
    title = "{Radiative Force from Optical Cycling on a Diatomic Molecule}",
  journal = {Phys. Rev. Lett.},
archivePrefix = "arXiv",
 primaryClass = "physics.atom-ph",
     year = 2009,
    month = nov,
   volume = 103,
   number = 22,
    pages = {223001},
      doi = {10.1103/PhysRevLett.103.223001},
   adsurl = {http://adsabs.harvard.edu/abs/2009PhRvL.103v3001S}
}

@ARTICLE{Stuhl2008,
   author = {{Stuhl}, B.~K. and {Sawyer}, B.~C. and {Wang}, D. and {Ye}, J.
	},
    title = "{Magneto-optical Trap for Polar Molecules}",
  journal = {Phys. Rev. Lett.},
archivePrefix = "arXiv",
 primaryClass = "physics.atom-ph",
     year = 2008,
    month = dec,
   volume = 101,
   number = 24,
    pages = {243002},
      doi = {10.1103/PhysRevLett.101.243002},
   adsurl = {http://adsabs.harvard.edu/abs/2008PhRvL.101x3002S}
}

@article{Carr2009,
  author={L.D. Carr and D. DeMille and R.V. Krems and J. Ye},
  title={Cold and ultracold molecules: science, technology and applications},
  journal={New J. Phys.},
  volume={11},
  number={5},
  pages={055049},
  year = {2009},
  doi = {10.1088/1367-2630/11/5/055049},
}

@article{Kloter2008,
author = {B. Kloter and C. Weber and D. Haubrich and D. Meschede and H. Metcalf},
collaboration = {},
title = {Laser cooling of an indium atomic beam enabled by magnetic fields},
publisher = {APS},
year = {2008},
journal = {Phys. Rev. A},
volume = {77},
number = {3},
eid = {033402},
pages = {033402},
}

@Article{Berkeland2002,
  title = {Destabilization of dark states and optical spectroscopy in Zeeman-degenerate atomic systems},
  author = {Berkeland, D. J. and Boshier, M. G.},
  journal = {Phys. Rev. A},
  volume = {65},
  number = {3},
  pages = {033413},
  numpages = {13},
  year = {2002},
  doi = "10.1103/PhysRevA.65.033413",
  publisher = {American Physical Society},
}

@article{McCarron2018,
author = {D. J. McCarron},
title = {Laser cooling and trapping molecules},
year = {2018},
journal = {J. Phys. B: At. Mol. Opt. Phys.},
volume = {51},
pages = {212001}
}

@book{Field04,
  author={H. Lefebvre-Brion and R.W. Field},
  title="The spectra and dynamics of diatomic molecules",
  year=2004,
  publisher={Elsevier},
  address="New York"
}

@book{Brown2003,
  title = {Rotational spectroscopy of diatomic molecules},
  publisher = {Cambridge Univ. Press},
  year = {2003},
  author = {J. M. Brown and A. Carrington},
  }

@BOOK{Budker2008,
  title = {Atomic Physics: An Exploration Through Problems and Solutions, 2nd. Ed.},
  publisher = {Oxford Univ. Press},
  year = {2008},
  author = {D. Budker and D. F. Kimball and D. P. DeMille}
}

@article{DiRosa2004,
   author = {{Di Rosa}, M.~D.},
    title = "{Laser-cooling molecules}",
  journal = {European Physical Journal D},
     year = 2004,
    month = nov,
   volume = 31,
    pages = {395-402},
      doi = {10.1140/epjd/e2004-00167-2},
   adsurl = {http://adsabs.harvard.edu/abs/2004EPJD...31..395D}
}

@article{Liu2017,
  title = {Magnetic Trapping of Cold Methyl Radicals},
  author = {Liu, Yang and Vashishta, Manish and Djuricanin, Pavle and Zhou, Sida and Zhong, Wei and Mittertreiner, Tony and Carty, David and Momose, Takamasa},
  journal = {Phys. Rev. Lett.},
  volume = {118},
  issue = {9},
  pages = {093201},
  numpages = {5},
  year = {2017},
  month = {Feb},
  publisher = {American Physical Society},
  doi = {10.1103/PhysRevLett.118.093201},
}

@article{Prehn2016,
  title = {Optoelectrical Cooling of Polar Molecules to Submillikelvin Temperatures},
  author = {Prehn, Alexander and Ibr\"ugger, Martin and Gl\"ockner, Rosa and Rempe, Gerhard and Zeppenfeld, Martin},
  journal = {Phys. Rev. Lett.},
  volume = {116},
  issue = {6},
  pages = {063005},
  numpages = {6},
  year = {2016},
  month = {Feb},
  publisher = {American Physical Society},
  doi = {10.1103/PhysRevLett.116.063005},
}

@article{Williams2017,
  author={H J Williams and S Truppe and M Hambach and L Caldwell and N J Fitch and E A Hinds and B E Sauer and M R Tarbutt},
  title={Characteristics of a magneto-optical trap of molecules},
  journal={New Journal of Physics},
  volume={19},
  number={11},
  pages={113035},
  year={2017},
}

@article{Hutzler2011b,
author = {Hutzler, Nicholas R. and Lu, Hsin-I and Doyle, John M.},
title = {The Buffer Gas Beam: An Intense, Cold, and Slow Source for Atoms and Molecules},
journal = {Chemical Reviews},
volume = {112},
number = {9},
pages = {4803-4827},
year = {2012},
doi = {10.1021/cr200362u}
}

@Article{Barry2011,
author = {Barry, J. F. and Shuman, E. S. and DeMille, D.},
title  = {A bright, slow cryogenic molecular beam source for free radicals},
journal  = {Phys. Chem. Chem. Phys.},
year  = {2011},
volume  = {13},
issue  = {42},
pages  = {18936--18947},
doi = {10.1039/C1CP20335E},
}

@Article{Ni08,
 title = {A High Phase-Space-Density Gas of Polar Molecules},
 author = {K.-K. Ni and S. Ospelkaus and M.H.G. de Miranda and A. Pe'er and
 B. Neyenhuis and J.J. Zirbel and S. Kotochigova and P.S. Julienne and D.S. Jin
 and J. Ye},
 journal = {Science},
 volume = {322},
 pages = {231},
 year = {2008},
}

@article{Chieda2011,
  volume = {84},
  journal = {Phys. Rev. A},
  month = dec,
  numpages = {10},
  author = {Chieda, M. A. and Eyler, E. E.},
  title = {Prospects for rapid deceleration of small molecules by optical bichromatic forces},
  year = {2011},
  issue = {6},
  publisher = {American Physical Society},
  pages = {063401},
  doi = {10.1103/PhysRevA.84.063401}
}

@article{Collopy2018,
  title = {3D Magneto-Optical Trap of Yttrium Monoxide},
  author = {Collopy, Alejandra L. and Ding, Shiqian and Wu, Yewei and Finneran, Ian A. and Anderegg, Lo\"{\i}c and Augenbraun, Benjamin L. and Doyle, John M. and Ye, Jun},
  journal = {Phys. Rev. Lett.},
  volume = {121},
  issue = {21},
  pages = {213201},
  numpages = {5},
  year = {2018},
  month = {Nov},
  publisher = {American Physical Society},
  doi = {10.1103/PhysRevLett.121.213201},
}

@article{Anderegg2017,
  title = {Radio Frequency Magneto-Optical Trapping of CaF with High Density},
  author = {Anderegg, Lo\"{\i}c and Augenbraun, Benjamin L. and Chae, Eunmi and Hemmerling, Boerge and Hutzler, Nicholas R. and Ravi, Aakash and Collopy, Alejandra and Ye, Jun and Ketterle, Wolfgang and Doyle, John M.},
  journal = {Phys. Rev. Lett.},
  volume = {119},
  issue = {10},
  pages = {103201},
  numpages = {5},
  year = {2017},
  month = {Sep},
  publisher = {American Physical Society},
  doi = {10.1103/PhysRevLett.119.103201},
}

@ARTICLE{Shuman2010,
   author = {{Shuman}, E.~S. and {Barry}, J.~F. and {DeMille}, D.},
   title = "{Laser cooling of a diatomic molecule}",
   journal =      {Nature},
   year =     {2010},
   volume =   {467},
   pages =    {820--823},
   doi = {10.1038/nature09443}
}

@article{Fitch2016,
author = {Fitch, N. J. and Tarbutt, M. R.},
title = {Principles and Design of a Zeeman–Sisyphus Decelerator for Molecular Beams},
journal = {ChemPhysChem},
volume = {17},
number = {22},
pages = {3609-3623},
doi = {10.1002/cphc.201600656},
year = {2016}
}

@article{Willitsch2018,
author = {Germann,Matthias  and Willitsch,Stefan },
title = {Fine- and hyperfine-structure effects in molecular photoionization. I. General theory and direct photoionization},
journal = {The Journal of Chemical Physics},
volume = {145},
number = {4},
pages = {044314},
year = {2016},
doi = {10.1063/1.4955301},


}

@article{ZACHWIEJA1995,
title = {New Investigations of the $A^{2}\Delta - X^{2}\Pi$  Band System in the CH Radical and a New Reduction of the Vibration-Rotation Spectrum of CH from the ATMOS Spectra},
journal = {Journal of Molecular Spectroscopy},
volume = {170},
number = {2},
pages = {285-309},
year = {1995},
issn = {0022-2852},
doi = {https://doi.org/10.1006/jmsp.1995.1072},
author = {M. Zachwieja},
}

@article{KEPA1996,
title = {New Spectroscopic Analysis of the $B^{2}\Sigma - X^{2}\Pi$ Band System of the CH Molecule},
journal = {Journal of Molecular Spectroscopy},
volume = {178},
number = {2},
pages = {189-193},
year = {1996},
issn = {0022-2852},
doi = {https://doi.org/10.1006/jmsp.1996.0173},
author = {R. Kepa and A. Para and M. Rytel and M. Zachwieja},
}

@article{LEROY2017,
title = {RKR1: A computer program implementing the first-order RKR method for determining diatomic molecule potential energy functions},
journal = {Journal of Quantitative Spectroscopy and Radiative Transfer},
volume = {186},
pages = {158-166},
year = {2017},
note = {Satellite Remote Sensing and Spectroscopy: Joint ACE-Odin Meeting, October 2015},
issn = {0022-4073},
doi = {https://doi.org/10.1016/j.jqsrt.2016.03.030},
author = {Robert J. {Le Roy}}
}

@Article{Lane2011,
author ={Wells, Nathan and Lane, Ian C.},
title  ={Prospects for ultracold carbon via charge exchange reactions and laser cooled carbides},
journal  ={Phys. Chem. Chem. Phys.},
year  ={2011},
volume  ={13},
issue  ={42},
pages  ={19036-19051},
publisher  ={The Royal Society of Chemistry},
doi  ={10.1039/C1CP21304K}
}

@ARTICLE{Lutz1970,
       author = {{Hesser}, James E. and {Lutz}, Barry L.},
        title = {Probabilities for Radiation and Predissociation. II. The Excited States of CH, CD, and $\rm{CH^{+}}$, and Some Astrophysical Implications},
      journal = {Am. Phys. Journal},
         year = {1970},
        month = {feb},
       volume = {159},
        pages = {703},
          doi = {10.1086/150344},

}

@article{Masseron2014,
	author = {{Masseron, T.} and {Plez, B.} and {Van Eck, S.} and {Colin, R.} and {Daoutidis, I.} and {Godefroid, M.} and {Coheur, P.-F.} and {Bernath, P.} and {Jorissen, A.} and {Christlieb, N.}},
	title = {CH in stellar atmospheres: an extensive linelist},
	DOI= "10.1051/0004-6361/201423956",
	journal = {Astronomy \& Astrophysics},
	year = 2014,
	volume = 571,
	pages = "A47",
	month = "",
}

@article{Kalemos1999,
author = {Kalemos,Apostolos  and Mavridis,Aristides  and Metropoulos,Aristophanes },
title = {An accurate description of the ground and excited states of CH},
journal = {The Journal of Chemical Physics},
volume = {111},
number = {21},
pages = {9536-9548},
year = {1999},
doi = {10.1063/1.480285},

}

@article{BERNATH1991,
title = {Spectroscopy of the CH free radical},
journal = {Journal of Molecular Spectroscopy},
volume = {147},
number = {1},
pages = {16-26},
year = {1991},
issn = {0022-2852},
doi = {https://doi.org/10.1016/0022-2852(91)90164-6},
author = {P.F. Bernath and C.R. Brazier and T. Olsen and R. Hailey and W.T.M.L. Fernando and Christine Woods and J.L. Hardwick}
}

@article{Cui2018,
	doi = {10.1088/1674-1056/27/10/103101},
	year = 2018,
	month = {oct},
	publisher = {{IOP} Publishing},
	volume = {27},
	number = {10},
	pages = {103101},
	author = {Jie Cui and Jian-Gang Xu and Jian-Xia Qi and Ge Dou and Yun-Guang Zhang},
	title = {Laser cooling of {CH} molecule: Insights from
		                    ab initio
		                    study},
	journal = {Chinese Physics B}
}

@article{NEMES1999,
author = { {P. G.   Szalay}  and  {L.   Nemes} },
title = {Tunnelling lifetimes of the rovibronic levels in the B electronic state of the CH radical obtained from ab initio data},
journal = {Molecular Physics},
volume = {96},
number = {3},
pages = {359-366},
year  = {1999},
publisher = {Taylor & Francis},
doi = {10.1080/00268979909482969},


}

@ARTICLE{Erman1976,
       author = {{Brzozowski}, J. and {Bunker}, P. and {Elander}, N. and {Erman}, P.},
        title = {Predissociation effects in the A, B and C states of CH and the interstellar formation rate of CH via inverse predissociation},
      journal = {Am. Phys. Journal},
         year = 1976,
        month = jul,
       volume = {207},
        pages = {414-424},
          doi = {10.1086/154509}
}

@article{Lancaster2020,
  title = {Electrostatic guiding of the methylidyne radical at cryogenic temperatures},
  author = {Lancaster, David M. and Allen, Cameron H. and Jersey, Kylan and Lancaster, Thomas A. and Shaw, Gage and Taylor, Mckenzie J. and Xiao, Di and Weinstein, Jonathan D.},
  journal = {The European Physical Journal D},
  volume = {74},
  issue = {6},
  pages = {132},
  year = {2020},
  month = {Jun},
  doi = {10.1140/epjd/e2020-10240-3},
}

@article{Fabrikant2014,
  title = {Method for traveling-wave deceleration of buffer-gas beams of CH},
  author = {Fabrikant, M. I. and Li, Tian and Fitch, N. J. and Farrow, N. and Weinstein, Jonathan D. and Lewandowski, H. J.},
  journal = {Phys. Rev. A},
  volume = {90},
  issue = {3},
  pages = {033418},
  numpages = {9},
  year = {2014},
  month = {Sep},
  publisher = {American Physical Society},
  doi = {10.1103/PhysRevA.90.033418},
}

@article{Shaw2020,
  title = {Bright, continuous beams of cold free radicals},
  author = {Shaw, J. C. and McCarron, D. J.},
  journal = {Phys. Rev. A},
  volume = {102},
  issue = {4},
  pages = {041302},
  numpages = {5},
  year = {2020},
  month = {Oct},
  publisher = {American Physical Society},
  doi = {10.1103/PhysRevA.102.041302},
}

@ARTICLE{Douglas1941,
       author = {{Douglas}, A.~E. and {Herzberg}, G.},
        title = "{Note on $\rm{CH^{+}}$ in Interstellar Space and in the Laboratory.}",
      journal = {The Astrophysical Journal},
         year = 1941,
        month = sep,
       volume = {94},
        pages = {381},
          doi = {10.1086/144342}
}

@article{Smith2011,
author = {Smith, Ian W.M.},
title = {Laboratory Astrochemistry: Gas-Phase Processes},
journal = {Annual Review of Astronomy and Astrophysics},
volume = {49},
number = {1},
pages = {29-66},
year = {2011},
doi = {10.1146/annurev-astro-081710-102533},

}

@article{Maksyutenko2011,
    author ={Maksyutenko, Pavlo and Zhang, Fangtong and Gu, Xibin and Kaiser, Ralf I.},
    title  ={A crossed molecular beam study on the reaction of methylidyne radicals [CH($\rm{X_{2}\Pi}$)] with acetylene [$\rm{C_{2}H_{2}(X^{1}\Sigma^{g+})}$]—competing $\rm{C_{3}H_{2}}$ + $H$ and $\rm{C_{3}H + H_{2}}$ channels},
    journal  = {Phys. Chem. Chem. Phys.},
    year  ={2011},
    volume  ={13},
    issue  ={1},
    pages  ={240-252},
    publisher  ={The Royal Society of Chemistry},
    doi  ={10.1039/C0CP01529F},
    }

@article {DeVine2017,
	author = {DeVine, Jessalyn A. and Weichman, Marissa L. and Laws, Benjamin and Chang, Jing and Babin, Mark C. and Balerdi, Garikoitz and Xie, Changjian and Malbon, Christopher L. and Lineberger, W. Carl and Yarkony, David R. and Field, Robert W. and Gibson, Stephen T. and Ma, Jianyi and Guo, Hua and Neumark, Daniel M.},
	title = {Encoding of vinylidene isomerization in its anion photoelectron spectrum},
	volume = {358},
	number = {6361},
	pages = {336--339},
	year = {2017},
	doi = {10.1126/science.aao1905},
	publisher = {American Association for the Advancement of Science},
	issn = {0036-8075},
	journal = {Science}
}

@article {Beyer2017,
	author = {Beyer, Axel and Maisenbacher, Lothar and Matveev, Arthur and Pohl, Randolf and Khabarova, Ksenia and Grinin, Alexey and Lamour, Tobias and Yost, Dylan C. and H{\"a}nsch, Theodor W. and Kolachevsky, Nikolai and Udem, Thomas},
	title = {The Rydberg constant and proton size from atomic hydrogen},
	volume = {358},
	number = {6359},
	pages = {79--85},
	year = {2017},
	doi = {10.1126/science.aah6677},
	publisher = {American Association for the Advancement of Science},
	issn = {0036-8075},
	journal = {Science}
}

@article{Yarkony1994,
author = {Hettema,Hinne  and Yarkony,David R. },
title = {On the radiative lifetime of the ($\rm{a^{4}\Sigma^{−},v,N,Fi}$) levels of the CH radical: An ab initio treatment},
journal = {The Journal of Chemical Physics},
volume = {100},
number = {12},
pages = {8991-8998},
year = {1994},
doi = {10.1063/1.466703},


    

}

@article{Augenbraun2020,
  title = {Molecular Asymmetry and Optical Cycling: Laser Cooling Asymmetric Top Molecules},
  author = {Augenbraun, Benjamin L. and Doyle, John M. and Zelevinsky, Tanya and Kozyryev, Ivan},
  journal = {Phys. Rev. X},
  volume = {10},
  issue = {3},
  pages = {031022},
  numpages = {20},
  year = {2020},
  month = {Jul},
  publisher = {American Physical Society},
  doi = {10.1103/PhysRevX.10.031022},
}

@article{vanDishoeck1987,
author = {van Dishoeck,Ewine F. },
title = {Photodissociation processes in the CH molecule},
journal = {The Journal of Chemical Physics},
volume = {86},
number = {1},
pages = {196-214},
year = {1987},
doi = {10.1063/1.452610},
    
}

@Article{Liu2020,
author ="Liu, Yu and Grimes, David D. and Hu, Ming-Guang and Ni, Kang-Kuen",
title  ="Probing ultracold chemistry using ion spectrometry",
journal  ="Phys. Chem. Chem. Phys.",
year  ="2020",
volume  ="22",
issue  ="9",
pages  ="4861-4874",
publisher  ="The Royal Society of Chemistry",
doi  ="10.1039/C9CP07015J"}

@article{Koslov2009,
  title = {$\ensuremath{\Lambda}$-doublet spectra of diatomic radicals and their dependence on fundamental constants},
  author = {Kozlov, M. G.},
  journal = {Phys. Rev. A},
  volume = {80},
  issue = {2},
  pages = {022118},
  numpages = {10},
  year = {2009},
  month = {Aug},
  publisher = {American Physical Society},
  doi = {10.1103/PhysRevA.80.022118},
}

@article {Mitra2020,
	author = {Mitra, Debayan and Vilas, Nathaniel B. and Hallas, Christian and Anderegg, Lo{\"\i}c and Augenbraun, Benjamin L. and Baum, Louis and Miller, Calder and Raval, Shivam and Doyle, John M.},
	title = {Direct laser cooling of a symmetric top molecule},
	volume = {369},
	number = {6509},
	pages = {1366--1369},
	year = {2020},
	doi = {10.1126/science.abc5357},
	publisher = {American Association for the Advancement of Science},
	issn = {0036-8075},
	journal = {Science}
}

@article{Umbachs1986,
author = {Ubachs,Wim  and Meyer,Gerard  and ter Meulen,J. J.  and Dymanus,A. },
title = {Hyperfine structure and lifetime of the $\rm{C^{2}\Sigma^{+}}$, v=0 state of CH},
journal = {The Journal of Chemical Physics},
volume = {84},
number = {6},
pages = {3032-3041},
year = {1986},
doi = {10.1063/1.450284},





}

@PhdThesis{TruppeThesis,
           title = "{New Physics with Cold Molecules: Precise Microwave Spectroscopy of CH and the Development of a Microwave Trap}",
          author = "Stefan Truppe",
          school = "Imperial College London",
            year = "2014",
            
}

@article{Augenbraun_2020,
	doi = {10.1088/1367-2630/ab687b},
	year = 2020,
	month = {feb},
	publisher = {{IOP} Publishing},
	volume = {22},
	number = {2},
	pages = {022003},
	author = {Benjamin L Augenbraun and Zack D Lasner and Alexander Frenett and Hiromitsu Sawaoka and Calder Miller and Timothy C Steimle and John M Doyle},
	title = {Laser-cooled polyatomic molecules for improved electron electric dipole moment searches},
	journal = {New Journal of Physics},
}

@article{Cooke2019,
author = {Cooke, Ilsa R. and Sims, Ian R.},
title = {Experimental Studies of Gas-Phase Reactivity in Relation to Complex Organic Molecules in Star-Forming Regions},
journal = {ACS Earth and Space Chemistry},
volume = {3},
number = {7},
pages = {1109-1134},
year = {2019},
doi = {10.1021/acsearthspacechem.9b00064},

}

@article{Burkley19,
author = {Z. Burkley and A. D. Brandt and C. Rasor and S. F. Cooper and D. C. Yost},
journal = {Appl. Opt.},
number = {7},
pages = {1657--1661},
publisher = {OSA},
title = {Highly coherent, watt-level deep-UV radiation via a frequency-quadrupled Yb-fiber laser system},
volume = {58},
month = {Mar},
year = {2019},
doi = {10.1364/AO.58.001657},
}

@article{Kielpinski2006,
  title = {Laser cooling of atoms and molecules with ultrafast pulses},
  author = {Kielpinski, D.},
  journal = {Phys. Rev. A},
  volume = {73},
  issue = {6},
  pages = {063407},
  numpages = {6},
  year = {2006},
  month = {Jun},
  publisher = {American Physical Society},
  doi = {10.1103/PhysRevA.73.063407},
}

@article{Baum2021,
  title = {Establishing a nearly closed cycling transition in a polyatomic molecule},
  author = {Baum, Louis and Vilas, Nathaniel B. and Hallas, Christian and Augenbraun, Benjamin L. and Raval, Shivam and Mitra, Debayan and Doyle, John M.},
  journal = {Phys. Rev. A},
  volume = {103},
  issue = {4},
  pages = {043111},
  numpages = {13},
  year = {2021},
  month = {Apr},
  publisher = {American Physical Society},
  doi = {10.1103/PhysRevA.103.043111},
}

@misc{Augenbraun2021,
      title={Zeeman-Sisyphus Deceleration of Molecular Beams}, 
      author={Benjamin L. Augenbraun and Alexander Frenett and Hiromitsu Sawaoka and Christian Hallas and Nathaniel B. Vilas and Abdullah Nasir and Zack D. Lasner and John M. Doyle},
      year={2021},
      eprint={2109.03067},
      archivePrefix={arXiv},
      primaryClass={physics.atom-ph}
}

\end{document}